\documentclass[conference]{IEEEtran}
\IEEEoverridecommandlockouts

\usepackage{cite}
\usepackage{amsmath,amssymb,amsfonts}
\usepackage{algorithmic}
\usepackage{graphicx}
\usepackage{textcomp}
\usepackage{xcolor}
\usepackage{subcaption}
\usepackage{tikz}
\usetikzlibrary{quantikz2}

\def\BibTeX{{\rm B\kern-.05em{\sc i\kern-.025em b}\kern-.08em
    T\kern-.1667em\lower.7ex\hbox{E}\kern-.125emX}}
\begin{document}

\title{Quantum Software Security Challenges within Shared Quantum Computing Environments\\

\thanks{This paper has been accepted for presentation at the 2025 IEEE International Conference on Quantum Computing and Engineering (QCE).}

\thanks{This work has been supported by Business Finland (Project: Securing the Quantum Software Stack (SeQuSoS) 112/31/2024).}
}

 \author{\IEEEauthorblockN{Samuel Ovaskainen}
 \IEEEauthorblockA{\textit{Faculty of Information Technology} \\
 \textit{University of Jyväskylä}\\
\textit{Jyväskylä, Finland} \\
 samuel.s.ovaskainen@student.jyu.fi}
 \and
 \IEEEauthorblockN{Majid Haghparast}
 \IEEEauthorblockA{\textit{Faculty of Information Technology} \\
 \textit{University of Jyväskylä}\\
\textit{Jyväskylä, Finland} \\
 majid.m.haghparast@jyu.fi}
 \and
 \IEEEauthorblockN{Tommi Mikkonen}
 \IEEEauthorblockA{\textit{Faculty of Information Technology} \\
 \textit{University of Jyväskylä}\\
\textit{Jyväskylä, Finland} \\
tommi.j.mikkonen@jyu.fi}
}

\maketitle

\begin{abstract}
The number of qubits in quantum computers keeps growing, but most quantum programs remain relatively small because of the noisy nature of the underlying quantum hardware. This might lead quantum cloud providers to explore increased hardware utilization, and thus profitability through means such as multi-programming, which would allow the execution of multiple programs in parallel. The adoption of such technology would bring entirely new challenges to the field of quantum software security. This article explores and reports the key challenges identified in quantum software security within shared quantum computing environments.
\end{abstract}


\section{Introduction} \label{section:introduction}
As the number of available qubits continues to increase in quantum processing units, the world continues its journey toward practical quantum software. In the future, we will see enormous benefits from quantum in terms of faster and more capable applications, potentially even allowing us to solve whole new classes of problems currently deemed unsolvable by classical computers. As our current understanding may only scrape the surface of the full potential realized by quantum computing, we are left with theories and educated guesses about its impact on our lives.

However, not everyone wants to play by the same rules, and new opportunities are rarely used only for good. Not only do quantum computers pose a risk to existing encryption methods readily used in software development, but as the field of quantum software grows, so will the number of attacks attempting to specifically target it. From nation-state adversaries to single bad actors, quantum software will face many of the same issues already faced by classical computing, and a whole new category of attacks attempting to specifically exploit the weaknesses of the emerging technology. 

In this paper, we survey some of the challenges in software security that the scientific community has identified for quantum computing, with a focus on shared computing environments. In addition, further research directions are presented, which we consider to be of particular importance for ensuring the security of quantum software. We believe these issues should be addressed before confidential data can be processed using shared quantum computing environments.

The remainder of this paper is organized as follows: Section~\ref{section:background} presents the background and motivation for this work. Section~\ref{section:multi} outlines key challenges in quantum software security, with a focus on multi-tenant computing. Section~\ref{section:calltoaction} provides a call to action for the research community. Finally, Section~\ref{section:conclusions} concludes the paper.

\section{Background and Motivation} \label{section:background}

Term \textbf{\textit{software security}} is often used without exact definitions. In the absence of an explicit definition, the term can mean slightly different things. In this paper, software security refers to the idea of designing software in a way that makes it resilient to adversarial attacks, protecting data to ensure its correctness and confidentiality~\cite{mcgraw_software_2004,gunnell_software_2024}

In cybersecurity, threat actors are often categorized with varying levels of specificity based on multiple characteristics including, but not limited to, motives and objectives. The threats highlighted for quantum computing are politically motivated \textit{nation-state actors}, financially motivated \textit{cybercriminals}, and ideologically motivated \textit{hacktivists}.

As present-day quantum computing is dominated by cloud-based approaches, the targets of threat actors are quantum application providers and their clients~\cite{hassija_present_2020}. Furthermore, because information is not transmitted through quantum channels, qubits are never sent directly to quantum computers. This means that all communication with quantum computer systems occurs through traditional computer systems, further expanding the attack surface~\cite{kilber_cybersecurity_2021}. However, these concerns fall outside the scope of this paper, and traditional vulnerabilities are better understood and documented.

\textbf{\textit{NISQ (Noisy Intermediate-Scale Quantum)}
}
    is a term coined by John Preskill and refers to the present-day quantum technology, which has a limited amount of qubits (50 - a few hundred) and only limited control over the state of the said qubits, imposing "noise" on the results and thus limiting the capabilities of the hardware~\cite{preskill_quantum_2018}.

\section{Quantum Software Security Challenges for Multi-tenant computing}\label{section:multi}
The large error rates in the present NISQ mean that only a limited number of gate operations can be performed while the results remain reliable. The limited scale of programs, in tandem with the growing amount of available qubits on quantum hardware means that in order to acquire better utilization rates multiple programs can be executed on the same hardware at the same time by means of multi-programming. Multi-programming is an approach to multi-tenant computing in which the qubits of the quantum computer are partitioned between multiple circuits, which are then executed in parallel~\cite{das_case_2019}. However, extra attention needs to be paid, as qubits have varying levels of connectivity and error rates, making fair resource allocation tricky. Such multi-programming approaches for improving utilization rates of quantum hardware have already been proposed~\cite{das_case_2019,niu_enabling_2023}.

As quantum computers require specialized conditions and equipment for the foreseeable future cloud-based quantum solutions such as \textit{serverless quantum} and \textit{quantum-as-a-service} are expected to remain the most popular options for companies and research institutes looking to leverage the benefits of quantum. Maximizing utilization and, in turn, the amount of paying customers served is in the best interest of for-profit cloud quantum computing service providers, which is why we can reasonably expect them to turn to multi-programming to increase profitability further. However, employing multi-programming in quantum computing has many unsolved challenges and vulnerabilities, which this paper will attempt to explore further below.

\subsection{Crosstalk}
Present-day NISQ machines suffer from errors caused by a subsystem (often a qubit or a control line) that unintentionally affect the behavior of another subsystem called \textit{crosstalk}. As the word crosstalk was borrowed from electrical engineering and its usage is imprecise, a precise quantum-specific definition \textit{crosstalk error} was coined. Crosstalk errors are defined as the behavior of quantum gates and circuits diverging from the ideal at the quantum logic level caused by physical crosstalk~\cite{sarovar_detecting_2020}. Crosstalk may even be the largest source of errors in quantum computers~\cite{rudinger_experimental_2021}. Despite the extremely high prevalence of crosstalk errors due to the large diversity in causes, its detailed characterization is said to be "extraordinarily difficult", and even simple detection can be cumbersome~\cite{sarovar_detecting_2020}. For these reasons, crosstalk is a very active field of research.

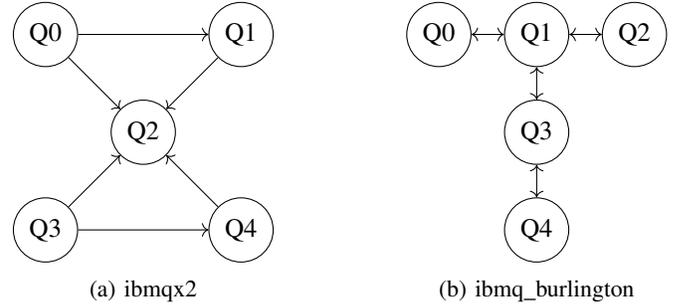
\begin{figure}[h]
    \begin{subfigure}[t]{0.2\textwidth}
        \centering
        \begin{tikzpicture}[scale=1.3]
            \node at (0,2) [circle, draw](Q0) {Q0};
            \node at (2,2) [circle, draw](Q1) {Q1};
            \node at (1,1) [circle, draw](Q2) {Q2};
            \node at (0,0) [circle, draw](Q3) {Q3};
            \node at (2,0) [circle, draw](Q4) {Q4};
            \path[->] (Q0) edge (Q1) edge (Q2)
            (Q1) edge (Q2)
            (Q4) edge (Q2)
            (Q3) edge (Q2) edge (Q4);
        \end{tikzpicture}
    \caption{ibmqx2}\label{ibmqx2}
    \end{subfigure}
    \hfill
    \begin{subfigure}[t]{0.2\textwidth}
        \centering
        \begin{tikzpicture}[scale=1.3]
            \node at (0,2) [circle, draw](Q0) {Q0};
            \node at (1,2) [circle, draw](Q1) {Q1};
            \node at (2,2) [circle, draw](Q2) {Q2};
            \node at (1,1) [circle, draw](Q3) {Q3};
            \node at (1,0) [circle, draw](Q4) {Q4};
            \path[<->] (Q0) edge (Q1)
            (Q1) edge (Q2) edge (Q3)
            (Q3) edge (Q4);
        \end{tikzpicture}
    \caption{ibmq\_burlington}\label{ibmqb}
    \end{subfigure}
    \caption{Topologies of two five-qubit IBM quantum computers}
\end{figure}

\begin{figure}[h]
    \begin{center}
    \begin{quantikz}
        \lstick{Q0 \ket 0} & \gate{X} & \ctrl{1}\gategroup[3, steps=2,style={inner sep=6pt, dashed, rounded corners, color=blue}]{\textcolor{blue}{Quantum half adder}} & \ctrl{1} && \\
        \lstick{Q1 \ket 0} & & \ctrl{1} & \targ{} && \\
        \lstick{Q2 \ket 0} && \targ{} &&& \\
        \lstick{Q3 \ket 0} & \ctrl[style={fill=red!,color=red!}]{1}\gategroup[2, steps=4,style={inner sep=6pt, dashed, rounded corners, color=red},label style={label
    position=below,anchor=north,yshift=-0.2cm}]{\textcolor{red}{Adversary CNOT gates}} & \ctrl[style={fill=red!,color=red!}]{1} &\ \ldots\ & \ctrl[style={fill=red!,color=red!}]{1} & \\
        \lstick{Q4 \ket 0} & \targ[style={color=red!}]{} & \targ[style={color=red!}]{} &\ \ldots\ & \targ[style={color=red!}]{} & 
    \end{quantikz}
    \end{center}
    \caption{Simplified version of the setup presented in~\cite{ash-saki_analysis_2020}.}\label{fqa}
\end{figure}
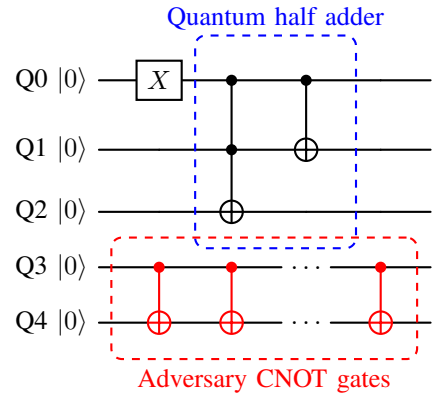

It has been stated that executing multiple programs in parallel, which is the objective of multi-programming, can further increase the prevalence of crosstalk errors~\cite{das_case_2019}. In addition to reducing the reliability of results under normal conditions, previous research has indicated that in present-day cloud-based quantum hardware, if multi-programming is utilized, crosstalk can be used by attackers for denial-of-service purposes or to influence the results of the victim in the shared computing environment. Ash Saki et al.~\cite{ash-saki_analysis_2020} established that the expected result state degrades proportionally to the number of adversary CNOT gates if the qubits in the circuits share a connection. Presented in Fig. \ref{ibmqx2} is the topology of ibmqx2, a five-qubit quantum computer, which due to its topology is stated to be vulnerable to the attack introduced in Fig. \ref{fqa}.

Using the introduced attack method on real-world IBM hardware, they found that running \textit{Grover-3 search algorithm} on the ibmqx2 would yield correct results less than $\frac 1 5$ of the time, becoming less likely than at least one wrong result after about 20 adversary CNOT gates. Similar results were replicated on ibmq\_burlington, the topology of which is presented in Fig. \ref{ibmqb}, where a wrong result became more likely after about 25 CNOT gates despite the differing layout~\cite{ash-saki_analysis_2020}. Similar attacks can be tailored for other hardware configurations if the attacker has a way of executing their program on qubits that share a connection to qubits utilized by the victim's program. 

As crosstalk is not only a problem between circuits but also a source of errors within them, reducing the overall reliability of quantum software, plenty of research has gone into approaches to mitigate it. Multiple strategies are explored in order to mitigate crosstalk, from hardware improvements to software-based approaches such as instruction scheduling, which attempts to reduce crosstalk by recognizing and rescheduling concurrent operations on crosstalk error-prone qubits~\cite{murali_software_2020}. Furthermore, crosstalk between circuits can be greatly reduced by allocating \textit{buffer qubits} between them to eliminate direct qubit connections between circuits, thus functioning as a primitive method of isolating circuits~\cite{ash-saki_analysis_2020}. A better understanding of the unwanted effects qubits have on each other will be crucial in creating mechanisms to adequately isolate quantum circuits.

\subsection{Adversarial SWAP Injection}
\begin{figure}[h]
    \begin{center}
    \begin{quantikz}[row sep={0.7cm, between origins}]
        && \swap{1} && \\
        && \targX{} && 
    \end{quantikz} \boldmath{=}
    \begin{quantikz}[row sep={0.7cm, between origins}, sep=0.4cm]
        & \ctrl{1} & \targ{} & \ctrl{1} & \\
        & \targ{} & \ctrl{-1} & \targ{} &
    \end{quantikz} 
    \end{center}
    \caption{SWAP gate implemented with 3 CNOT gates.}\label{swap}
\end{figure}
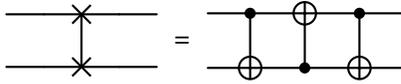

Adversarial SWAP injection is another attack that exploits the noisy nature and limited connections between qubits in present-day NISQ hardware in a multi-programming environment in order to increase the error rate in victim programs. The injection attack works with the principle of strategically occupying the most densely connected qubits in a quantum computer, which due to connectivity limitations forces the compiler to add additional SWAP gates to accommodate the victim circuit~\cite{upadhyay_stealthy_2023}. In large numbers, these additional SWAP gates -- as is currently the case with all gates -- decrease the reliability of the victim circuit. Furthermore, SWAP gates are often implemented using a number of other gates, further increasing the final number of gates, and, in turn, the rate of errors. An example of a SWAP gate implemented using 3 CNOT gates is presented in Fig.~\ref{swap}.

\begin{figure}[h]
    \begin{subfigure}[t]{0.23\textwidth}
        \centering
            \resizebox{\textwidth}{!}{
            \begin{quantikz}[row sep=0.3cm, sep=0.2cm, scale=0.4]
                \lstick{Q0} & \gate{H} & \ctrl{1} && \ctrl{1} & \swap{1} && \ctrl{3} & \\
                \lstick{Q1} & \gate{H} & \targ{} & \ctrl{1} & \targ{} & \targX{} &&& \\
                \lstick{Q2} & \gate{H} && \targ{} && \ctrl{1} &&& \\
                \lstick{Q3} & \gate{H} &&&& \targ{} && \targ{} & \\
            \end{quantikz}
        }
        \caption{Victim circuit running on qubits Q0 - Q3.}\label{swapv}
    \end{subfigure} \hfill
    \begin{subfigure}[t]{0.23\textwidth}
        \centering
         \resizebox{\textwidth}{!}{
            \begin{quantikz}[row sep=0.3cm, sep=0.2cm, scale=0.4, wire types={q, q, n, q, q}]
                \lstick{Q0} & \gate{H} & \ctrl{1} && \ctrl{1} & \swap{1} && \ctrl{4} & \\
                \lstick{Q1} & \gate{H} & \targ{} & \ctrl{2} & \targ{} & \targX{} &&& \\
                \lstick{\textcolor{red}{Q2}} &&&& \wire[l][4][style=red]{q}\ \textcolor{red}{\cdots}\ \wire[r][4][style=red]{q} &&&&  \\
                \lstick{Q3} & \gate{H} && \targ{} && \ctrl{1} & \swap[style=red]{1} && \\
                \lstick{Q4} & \gate{H} &&&& \targ{} & \targX[style=red]{} & \targ{} &
        \end{quantikz}
        }
        \caption{Victim circuit with an additional SWAP gate due to attacker occupying Q2.}\label{swapi}
    \end{subfigure}
    \caption{Victim program and adversarial swap injection presented on hardware equivalent to Fig. \ref{ibmqb}~\cite{upadhyay_stealthy_2023}.}\label{swapb}
\end{figure}
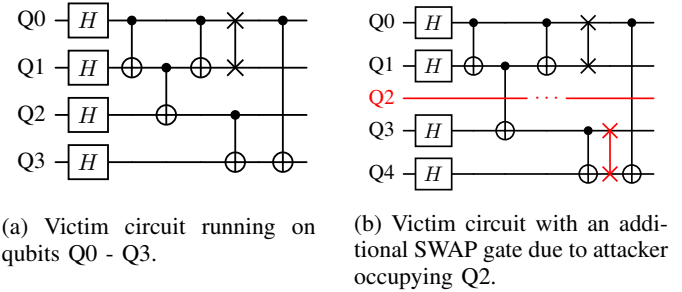

A simple demonstrative case of a victim circuit and an exploit running on quantum hardware with a layout identical to Fig. \ref{ibmqb} is presented in Fig. \ref{swapb}. In this case, by strategically occupying the Q2 qubit, the attacker forces the compiler to add an additional SWAP gate to the victim circuit. In more complex scenarios, it was demonstrated that on a 20 qubit simulator in some configurations a very significant amount of SWAP gates could be injected into the circuit: In the most effective configuration, a median increase of 25\% and a maximum increase of 55\% in SWAP gates were observed, when tested on 100 different victim circuits varying from 4 to 10 qubits in size. The second most effective configuration still showed a median increase of 20\%, and even the least efficient tested configuration displayed a median increase of 8\%. Logically, the total number of added SWAP gates increased with the length of the program, while the percentage of increase decreased as the program contained more gates to begin with. This attack was stated to be defended by employing anomaly detection in scheduling in order to detect possibly malicious patterns in task submissions~\cite{upadhyay_stealthy_2023}.

\subsection{Qubit sensing}
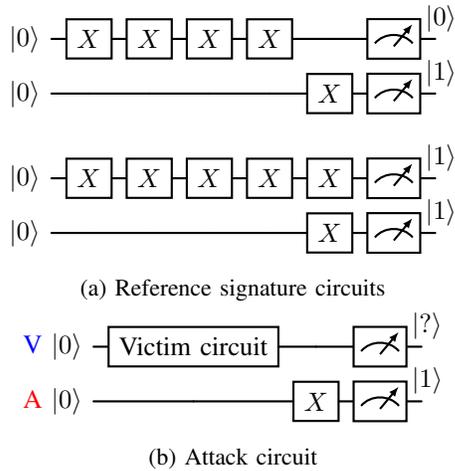
\begin{figure}[h]
    \begin{subfigure}[b]{0.5\textwidth}
        \centering
        \begin{quantikz}[row sep=0.3cm, sep=0.2cm, scale=0.4]
            \lstick{\ket{0}} & \gate{X}&\gate{X}&\gate{X}& \gate{X} && \meter[label style={right, xshift=0.3cm}]{\ket{0}} & \\
            \lstick{\ket{0}} &&&&& \gate{X} & \meter[label style={right, xshift=0.3cm}]{\ket{1}} & \\[0.4cm]
            \lstick{\ket{0}} & \gate{X}&\gate{X}&\gate{X}& \gate{X} & \gate{X} & \meter[label style={right, xshift=0.3cm}]{\ket{1}} & \\
            \lstick{\ket{0}} &&&&& \gate{X} & \meter[label style={right, xshift=0.3cm}]{\ket{1}} &
        \end{quantikz}
        \caption{Reference signature circuits}\label{qbst}
    \end{subfigure}
    \begin{subfigure}[b]{0.5\textwidth}
        \centering
        \begin{quantikz}[row sep=0.3cm, sep=0.2cm, scale=0.4]
            \lstick{\textcolor{blue}{V} \ket{0}} & \gate{\text{Victim circuit}} && \meter[label style={right, xshift=0.3cm}]{\ket{?}} & \\
            \lstick{\textcolor{red}{A} \ket{0}} && \gate{X} & \meter[label style={right, xshift=0.3cm}]{\ket{1}} &
        \end{quantikz}
        \caption{Attack circuit}\label{qbsa}
    \end{subfigure}
    \caption{Circuits utilized in qubit sensing attack~\cite{saki_qubit_2021}.}
\end{figure}

Qubit sensing is an attack model presented in ~\cite{saki_qubit_2021} that takes advantage of the noisy and unreliable nature of present-day quantum hardware where qubits have an unwanted influence on each other. The attack can be used to determine the output of a victim's circuit without having been granted access to the results in a multi-programming environment. Qubit sensing requires that the \textit{reference signature} of the hardware is known, which requires previous access to the hardware to conduct necessary measurements. The reference signature can be acquired through measurement using the measurement circuits presented in Fig. \ref{qbst}. 
After the attack circuit, presented in Fig. \ref{qbsa}, and the unknown victim circuit has been running in the same multi-programming environment, adversaries can attempt to identify the result of the victim circuit by analyzing the statistical distance between the unknown victim qubits and the reference signatures, as the probability of the adversary qubit is correlated with the victim qubit's result. When classifying an adjacent qubit the value was identified correctly up to 96\% of the time; however, it was stated that even non-adjacent qubits have an effect on each other and thus the attack is not limited to connected qubits. Furthermore, they found that one adversary qubit could be used to sense two victim qubits~\cite{saki_qubit_2021}.

Without additional attacks being used to find more information about the victim circuits or inputs, qubit sensing may have limited uses as it requires the context to be deciphered from only the result. On top of this, attackers need to know precisely on which qubits the circuits were executed. Additionally, there are some methods that can be used to mitigate qubit sensing attacks: The authors of the article proposed the simple measure of inserting NOT gates to invert some of the final output qubits, causing attackers to acquire an invalid output while the victim can correctly interpret the inverted qubits~\cite{saki_qubit_2021}. Furthermore, cloud providers could also attempt to obfuscate the machine and its topology from customers. However, this can potentially be circumvented using methods discussed in Section \ref{blueprinting}.

\subsection{Pulse-level attacks}\label{pulse-level}

Although quantum computers are commonly programmed with quantum gates, some quantum computer platforms give users the ability to execute control pulses, which is the method that quantum computers use under the hood to perform quantum gate operations. Control pulses give users more precise control over qubits as they enable the manipulation of the state in a continuous manner. Additionally, these control pulses allow qubits to reach states that cannot be reached using regular quantum gates~\cite{smith_pulselevel_2022}. However, the additional freedom given to users may allow new threats in a shared computing environment.

Attacks, called higher-energy state attacks, rely on these control pulses to reach energy states, such as $\ket{2}$, which are unexpected. Between runs, quantum computers use a so-called "reset" gate to initialize the qubits back to the ground state $\ket{0}$. However, it has been found that these reset gates, and in fact all gates, have almost no impact on these higher-energy qubits. As such, it is possible for adversaries to utilize these higher energy states in a multitude of ways to extract information or distort the results of victim circuits, assuming that the circuits are executed in an alternating fashion~\cite{xu_securing_2023}. However, right now this attack appears mostly theoretical as the delay caused by the loading of circuits on IBM machines is too long for these higher energy states not to collapse. Additionally, these load delays make running shots in an alternating fashion economically less lucrative, disincentivizing companies from exploring it as an option. The study assumes a future scenario in which these load delays have been reduced or completely eliminated.

Another pulse-level attack, called the QubitHammer attack, may induce severe crosstalk on a victim circuit even with a large buffer of qubits between the circuits in superconducting qubit quantum computers. The attack requires previous calibration on the machine in order to determine which frequency and amplitude have the highest impact on victim qubits. With enough adversary qubits, the attack has been shown to be highly effective in disrupting the results of the victim circuit. Additionally, the attack has been shown to bypass existing defense mechanisms against crosstalk, such as crosstalk-aware qubit allocation~\cite{tan2025qubithammer}.

Although a possible solution, completely deactivating pulse-level access may be undesirable, as pulse-level control is required in some valid applications such as quantum machine learning~\cite{tan2025qubithammer}. Thus, the mitigation of pulse-level attacks should be further investigated. These attacks highlight that buffer qubits alone may not be enough to isolate quantum circuits, which means that in a multi-tenant architecture, compilers may need to be aware of hardware limitations when scheduling the execution of quantum circuits.

\subsection{Circuit reconstruction}

As already demonstrated in Section \ref{pulse-level}, attacks in shared computing environments do not necessarily require quantum circuits to be executed simultaneously. Similarly, it has been found that it may be possible to extract information about a victim circuit if the execution queue is manipulated in a manner that allows the execution of a circuit before and after a victim circuit. By feeding a previously trained model the measurements of two ''probing'' circuits consisting of Hadamard gates, Bell and Trügler~\cite{bell_reconstructing_2022} established that a neural network was able to distinguish which of two possible test circuits had been executed between the circuits approximately 65\% of the time, demonstrating a potential source of information leakage.

However, it should be noted that the tested circuits consisted only of one quantum gate each. Additionally, the result was tested only on previously accessible five-qubit IBM quantum computers. The limited number of qubits greatly restricts the number of ways in which the gates could have been placed. As such, the authors presented the achieved result as a proof-of-concept, stating that the training data would have to be greatly expanded for a full side-channel attack~\cite{bell_reconstructing_2022}. The presented attack is simplistic in nature, but the results are still troublesome, as it highlights yet another attack model through which adversaries may be able to extract information in a shared computing environment. It is unclear how quantum service providers may prevent this kind of attack without reducing profitability by inducing long delays between circuit executions. 

\subsection{Hardware blueprinting}\label{blueprinting}
As many of the attacks discussed rely on the attacker knowing the topology of the quantum computer on which the circuit is running, concealing this information from end users may sound like a simple mitigation for a range of attacks. However, successful concealment may not be feasible as different methods for fingerprinting and identifying quantum hardware have already been developed. For example, crosstalk can be used with extremely high accuracy to determine on which quantum machine a circuit is running compared to previously acquired training data~\cite{mi_short_2022}. Similarly, it has been demonstrated that timing information could be used to determine the quantum processor used in only about 10 measurements~\cite{lu_quantum_2024}.

\section{Call to Action} \label{section:calltoaction}

To address the growing concerns surrounding quantum software security in shared quantum computing environments, we urge the research community, cloud providers, and policymakers to pursue the following directions:

\begin{itemize}
    \item \textbf{Development of Quantum Isolation Mechanisms:} Just as classical cloud computing evolved containerization, the quantum counterpart needs practical mechanisms to enforce execution isolation among concurrent quantum programs.
    
    \item \textbf{Security-Aware Quantum Compilers:} Designing quantum compilers that can reason about and enforce security constraints and account for identified vulnerabilities during resource allocation is crucial for safe and secure shared usage.
    
    \item \textbf{Quantum-Specific Side-Channel Mitigation:} Investigating and mitigating quantum-specific side-channel vulnerabilities such as crosstalk, QubitHammer, and adversarial SWAP injection in multi-tenant architectures is critical.

    \item \textbf{Benchmarking and Simulation Frameworks for Quantum Software Security Evaluation:} Building open-source tools to simulate quantum multi-tenancy and testing the effectiveness of different security mechanisms against realistic attack vectors is another research direction.
    
    \item \textbf{Hardware--Software Co-Design for Secure Quantum Architectures:} Promoting interdisciplinary collaboration to co-design quantum chips and control software that natively support secure resource partitioning and resistance to tampering needs further study.

    \item \textbf{Mitigation of Crosstalk Effects:} There is an urgent need to understand and mitigate crosstalk effects to secure current and future cloud-based quantum systems.

\end{itemize}

\section{Conclusion} \label{section:conclusions}
This paper provides insight into the emerging security challenges in quantum software and advocates for targeted research to address these concerns. Although most of the attacks presented in this paper were merely proofs of concept and had known countermeasures, the present situation already paints a stark picture for the future landscape of quantum computing: Not unlike in traditional computing, for countermeasures to work, we need to be aware of the exploits, which usually means discovering them after an incident has occurred. 

Quantum computing appears to be entering a continuous cycle of security challenges and countermeasures, reminiscent of the dynamics observed in classical computing over the past decades.
The similarities do not end there, and blueprints for \textit{quantum antiviruses} have already been proposed~\cite{deshpande_towards_2022,deshpande_design_2023}.
As quantum computers become more powerful and reliable, the size of programs executed on them will also grow in tandem. These larger programs will produce more interesting results, make it easier to conceal malicious circuits within them, and pose an ever more difficult challenge to defend against more sophisticated attacks. It remains to be seen whether quantum computing has an ace up its sleeves or if we are determined to repeat history by always being one step behind malicious actors. Perhaps new tools, such as artificial intelligence-based approaches, can be used to gain the upper hand.

\bibliographystyle{IEEEtran}
\bibliography{IEEEabrv,refs}

\begin{thebibliography}{10}
\providecommand{\url}[1]{#1}
\csname url@samestyle\endcsname
\providecommand{\newblock}{\relax}
\providecommand{\bibinfo}[2]{#2}
\providecommand{\BIBentrySTDinterwordspacing}{\spaceskip=0pt\relax}
\providecommand{\BIBentryALTinterwordstretchfactor}{4}
\providecommand{\BIBentryALTinterwordspacing}{\spaceskip=\fontdimen2\font plus
\BIBentryALTinterwordstretchfactor\fontdimen3\font minus \fontdimen4\font\relax}
\providecommand{\BIBforeignlanguage}[2]{{%
\expandafter\ifx\csname l@#1\endcsname\relax
\typeout{** WARNING: IEEEtran.bst: No hyphenation pattern has been}%
\typeout{** loaded for the language `#1'. Using the pattern for}%
\typeout{** the default language instead.}%
\else
\language=\csname l@#1\endcsname
\fi
#2}}
\providecommand{\BIBdecl}{\relax}
\BIBdecl

\bibitem{mcgraw_software_2004}
G.~McGraw, ``Software security,'' \emph{IEEE Security \& Privacy}, vol.~2, no.~2, pp. 80--83, Mar. 2004.

\bibitem{gunnell_software_2024}
\BIBentryALTinterwordspacing
M.~Gunnell, ``\BIBforeignlanguage{en-US}{Software {Security}},'' Jan. 2024. [Online]. Available: \url{https://www.techopedia.com/definition/24866/software-security}
\BIBentrySTDinterwordspacing

\bibitem{hassija_present_2020}
\BIBentryALTinterwordspacing
V.~Hassija, V.~Chamola, V.~Saxena, V.~Chanana, P.~Parashari, S.~Mumtaz, and M.~Guizani, ``Present landscape of quantum computing,'' vol.~1, no.~2, pp. 42--48, 2020. [Online]. Available: \url{https://onlinelibrary.wiley.com/doi/abs/10.1049/iet-qtc.2020.0027}
\BIBentrySTDinterwordspacing

\bibitem{kilber_cybersecurity_2021}
\BIBentryALTinterwordspacing
N.~Kilber, D.~Kaestle, and S.~Wagner, ``\BIBforeignlanguage{en}{Cybersecurity for {Quantum} {Computing}},'' Oct. 2021, arXiv:2110.14701 [quant-ph]. [Online]. Available: \url{http://arxiv.org/abs/2110.14701}
\BIBentrySTDinterwordspacing

\bibitem{preskill_quantum_2018}
J.~Preskill, ``\BIBforeignlanguage{en}{Quantum {Computing} in the {NISQ} era and beyond},'' \emph{\BIBforeignlanguage{en}{Quantum}}, vol.~2, p.~79, Aug. 2018.

\bibitem{das_case_2019}
P.~Das, S.~S. Tannu, P.~J. Nair, and M.~Qureshi, ``A {Case} for {Multi}-{Programming} {Quantum} {Computers},'' in \emph{Proceedings of the 52nd {Annual} {IEEE}/{ACM} {International} {Symposium} on {Microarchitecture}}, ser. {MICRO} '52.\hskip 1em plus 0.5em minus 0.4em\relax New York, NY, USA: Association for Computing Machinery, Oct. 2019, pp. 291--303.

\bibitem{niu_enabling_2023}
S.~Niu and A.~Todri-Sanial, ``\BIBforeignlanguage{en-GB}{Enabling {Multi}-programming {Mechanism} for {Quantum} {Computing} in the {NISQ} {Era}},'' \emph{\BIBforeignlanguage{en-GB}{Quantum}}, vol.~7, p. 925, Feb. 2023.

\bibitem{sarovar_detecting_2020}
M.~Sarovar, T.~Proctor, K.~Rudinger, K.~Young, E.~Nielsen, and R.~Blume-Kohout, ``\BIBforeignlanguage{en}{Detecting crosstalk errors in quantum information processors},'' \emph{\BIBforeignlanguage{en}{Quantum}}, vol.~4, p. 321, Sep. 2020, arXiv:1908.09855 [quant-ph].

\bibitem{rudinger_experimental_2021}
K.~Rudinger, C.~W. Hogle, R.~K. Naik, A.~Hashim, D.~Lobser, D.~I. Santiago, M.~D. Grace, E.~Nielsen, T.~Proctor, S.~Seritan, S.~M. Clark, R.~Blume-Kohout, I.~Siddiqi, and K.~C. Young, ``Experimental {Characterization} of {Crosstalk} {Errors} with {Simultaneous} {Gate} {Set} {Tomography},'' \emph{PRX Quantum}, vol.~2, no.~4, p. 040338, Nov. 2021, publisher: American Physical Society.

\bibitem{ash-saki_analysis_2020}
A.~Ash~Saki, M.~Alam, and S.~Ghosh, ``Analysis of crosstalk in {NISQ} devices and security implications in multi-programming regime,'' in \emph{Proceedings of the {ACM}/{IEEE} {International} {Symposium} on {Low} {Power} {Electronics} and {Design}}, ser. {ISLPED} '20.\hskip 1em plus 0.5em minus 0.4em\relax New York, NY, USA: Association for Computing Machinery, Aug. 2020, pp. 25--30.

\bibitem{murali_software_2020}
P.~Murali, D.~C. Mckay, M.~Martonosi, and A.~Javadi-Abhari, ``Software {Mitigation} of {Crosstalk} on {Noisy} {Intermediate}-{Scale} {Quantum} {Computers},'' in \emph{Proceedings of the {Twenty}-{Fifth} {International} {Conference} on {Architectural} {Support} for {Programming} {Languages} and {Operating} {Systems}}, ser. {ASPLOS} '20.\hskip 1em plus 0.5em minus 0.4em\relax New York, NY, USA: Association for Computing Machinery, Mar. 2020, pp. 1001--1016.

\bibitem{upadhyay_stealthy_2023}
S.~Upadhyay and S.~Ghosh, ``Stealthy {SWAPs}: {Adversarial} {SWAP} {Injection} in {Multi}-{Tenant} {Quantum} {Computing},'' Oct. 2023, arXiv:2310.17426 [quant-ph].

\bibitem{saki_qubit_2021}
\BIBentryALTinterwordspacing
A.~Ash~Saki and S.~Ghosh, ``\BIBforeignlanguage{en}{Qubit {Sensing}: {A} {New} {Attack} {Model} for {Multi}-programming {Quantum} {Computing}},'' Apr. 2021, arXiv:2104.05899 [quant-ph]. [Online]. Available: \url{http://arxiv.org/abs/2104.05899}
\BIBentrySTDinterwordspacing

\bibitem{smith_pulselevel_2022}
K.~N. Smith, G.~S. Ravi, T.~Alexander, N.~T. Bronn, A.~R.~R. Carvalho, A.~Cervera-Lierta, F.~T. Chong, J.~M. Chow, M.~Cubeddu, A.~Hashim, L.~Jiang, O.~Lanes, M.~J. Otten, D.~I. Schuster, P.~Gokhale, N.~Earnest, and A.~Galda, ``Programming physical quantum systems with pulse-level control,'' \emph{Frontiers in Physics}, vol. Volume 10 - 2022, 2022.

\bibitem{xu_securing_2023}
C.~Xu, J.~Chen, A.~Mi, and J.~Szefer, ``Securing {NISQ} {Quantum} {Computer} {Reset} {Operations} {Against} {Higher} {Energy} {State} {Attacks},'' in \emph{Proceedings of the 2023 {ACM} {SIGSAC} {Conference} on {Computer} and {Communications} {Security}}, ser. {CCS} '23.\hskip 1em plus 0.5em minus 0.4em\relax New York, NY, USA: Association for Computing Machinery, Nov. 2023, pp. 594--607.

\bibitem{tan2025qubithammer}
Y.~Tan, N.~Choudhury, K.~Basu, and J.~Szefer, ``Qubithammer attacks: Qubit flipping attacks in multi-tenant superconducting quantum computers,'' \emph{arXiv preprint arXiv:2504.07875}, 2025.

\bibitem{bell_reconstructing_2022}
B.~Bell and A.~Trügler, ``Reconstructing quantum circuits through side-channel information on cloud-based superconducting quantum computers,'' in \emph{2022 IEEE International Conference on Quantum Computing and Engineering (QCE)}, 2022, pp. 259--264.

\bibitem{mi_short_2022}
A.~Mi, S.~Deng, and J.~Szefer, ``Short {Paper}: {Device}- and {Locality}-{Specific} {Fingerprinting} of {Shared} {NISQ} {Quantum} {Computers},'' in \emph{Proceedings of the 10th {International} {Workshop} on {Hardware} and {Architectural} {Support} for {Security} and {Privacy}}, ser. {HASP} '21.\hskip 1em plus 0.5em minus 0.4em\relax New York, NY, USA: Association for Computing Machinery, Jun. 2022, pp. 1--6.

\bibitem{lu_quantum_2024}
C.~Lu, E.~Telang, A.~Aysu, and K.~Basu, ``Quantum {Leak}: {Timing} {Side}-{Channel} {Attacks} on {Cloud}-{Based} {Quantum} {Services},'' Jan. 2024, arXiv:2401.01521 [cs].

\bibitem{deshpande_towards_2022}
S.~Deshpande, C.~Xu, T.~Trochatos, Y.~Ding, and J.~Szefer, ``Towards an {Antivirus} for {Quantum} {Computers},'' in \emph{2022 {IEEE} {International} {Symposium} on {Hardware} {Oriented} {Security} and {Trust} ({HOST})}, Jun. 2022, pp. 37--40.

\bibitem{deshpande_design_2023}
S.~Deshpande, C.~Xu, T.~Trochatos, H.~Wang, F.~Erata, S.~Han, Y.~Ding, and J.~Szefer, ``Design of {Quantum} {Computer} {Antivirus},'' in \emph{2023 {IEEE} {International} {Symposium} on {Hardware} {Oriented} {Security} and {Trust} ({HOST})}, May 2023, pp. 260--270, iSSN: 2765-8406.

\end{thebibliography}

\end{document}